\documentclass[11pt]{article}
\usepackage{blois,epsfig}

\def\be{\begin{equation}}
\def\ee{\end{equation}}
\def\bea{\begin{eqnarray}}
\def\eea{\end{eqnarray}}

\begin{document}
\vspace*{4cm}
\title{COMPUTING NON-GAUSSIAN MAPS OF THE CMB}

\author{M. LANDRIAU~\footnote{Communicating author}}

\address{Laboratoire de l'Acc\'el\'erateur Lin\'eaire, IN2P3-CNRS et
Universit\'e Paris-Sud, B.P. 34, 91898 Orsay Cedex, France}

\author{E.P.S. SHELLARD}

\address{Department of Applied Mathematics and Theoretical Physics,
Centre for Mathematical Sciences, Wilberforce Road, Cambridge, CB3
0WA, United Kingdom}

\maketitle\abstracts{We discuss methods to compute maps of the CMB in
models featuring active causal sources and in non-Gaussian models of
inflation.  We show our large angle results as well as some
preliminary results on small angles.  We conclude by discussing
on-going work.}

\section{Introduction}

In the last few years, observations of the CMB, such as those from the
WMAP satellite~\cite{wmap1}, have provided increasingly strong evidence
for cosmic inflation, the idea that quantum fluctuations in the early
Universe were magnified up to become the primordial seeds for
structure formation.  In many models, inflation produces a purely
Gaussian signal.  In such a situation, it is straightforward to
compute a map, as all the information is contained in the two-point
correlation function, or equivalently in the $C_{\ell}$'s, which can
be computed in a few minutes at the most.  If we are interested in
producing only a temperature map, we only need to generate $2\ell +1$
random numbers with a Gaussian distribution of zero mean and a
standard deviation of $\sqrt{C_{\ell}}$ for every $\ell$ up to the
scale of interest and sum over the spherical harmonics.  In cases
where the signal is not purely Gaussian one cannot proceed in this
simple way as the $C_{\ell}$'s do not contain all the information.
The maps then have to be computed directly.

\section{Methods}

The method we will discuss here was developed to compute maps of the
CMB in cosmic string models and was presented in great detail
elsewhere~\cite{methods}.  Here, we shall go over the main points and show
how easily it can be adapted to treat non-Gaussian inflation.

All the metric and matter perturbation equations can be written in the
following manner:
\begin{equation}
\dot{y}(\mathbf{k},\eta) = \mathcal{A}(k,\eta) y(\mathbf{k},\eta)
+ q(\mathbf{k},\eta) \, ,
\end{equation}
with initial conditions $y(\mathbf{k},0) = c(\mathbf{k})$.  The vector
$q$ contains the causal source energy momentum tensor.  The solution
to such a coupled system of equation is:
\begin{equation}\label{eq:soln}
y(\mathbf{k},\eta) = Y(k,\eta)\mathbf{c(\mathbf{k},0)} 
+ Y(k,\eta)\int_0^{\eta}
Y^{-1}(k,\eta^{\prime}) q(\mathbf{k},\eta^{\prime}) d\eta^{\prime} \, ,
\end{equation}
where the fundamental matrix $Y$ obeys the following equation:
\begin{equation}
\dot{Y}(k,\eta) =  \mathcal{A}(k,\eta) Y(k,\eta) \, ,
\end{equation}
with initial conditions $Y(k,0) = \mathcal{I}$.  The matrices are
constructed from a Boltzmann code and are integrated over to produce
the perturbations which are integrated over according to the
Sachs-Wolfe formula to produce the temperature fluctuations.

\section{Cosmic Strings}

As mentionned in the previous section, we developed our methods to
treat cosmic strings.  Obviously, it can be used for other active
causal sources, but cosmic strings are inherently more interesting, as
they are expected to be generically produced at the end of inflation
in many models, notably many realistic versions of brane
inflation~\cite{sarangi2002}.

Our work involves Nambu-Goto strings, which are effectively described
as one-dimensional objects.  The equations of motions are solved using
the Allen-Shellard code~\cite{allen90}.

\subsection{All-Sky Maps}

\begin{figure}
\resizebox{77mm}{!}{\includegraphics{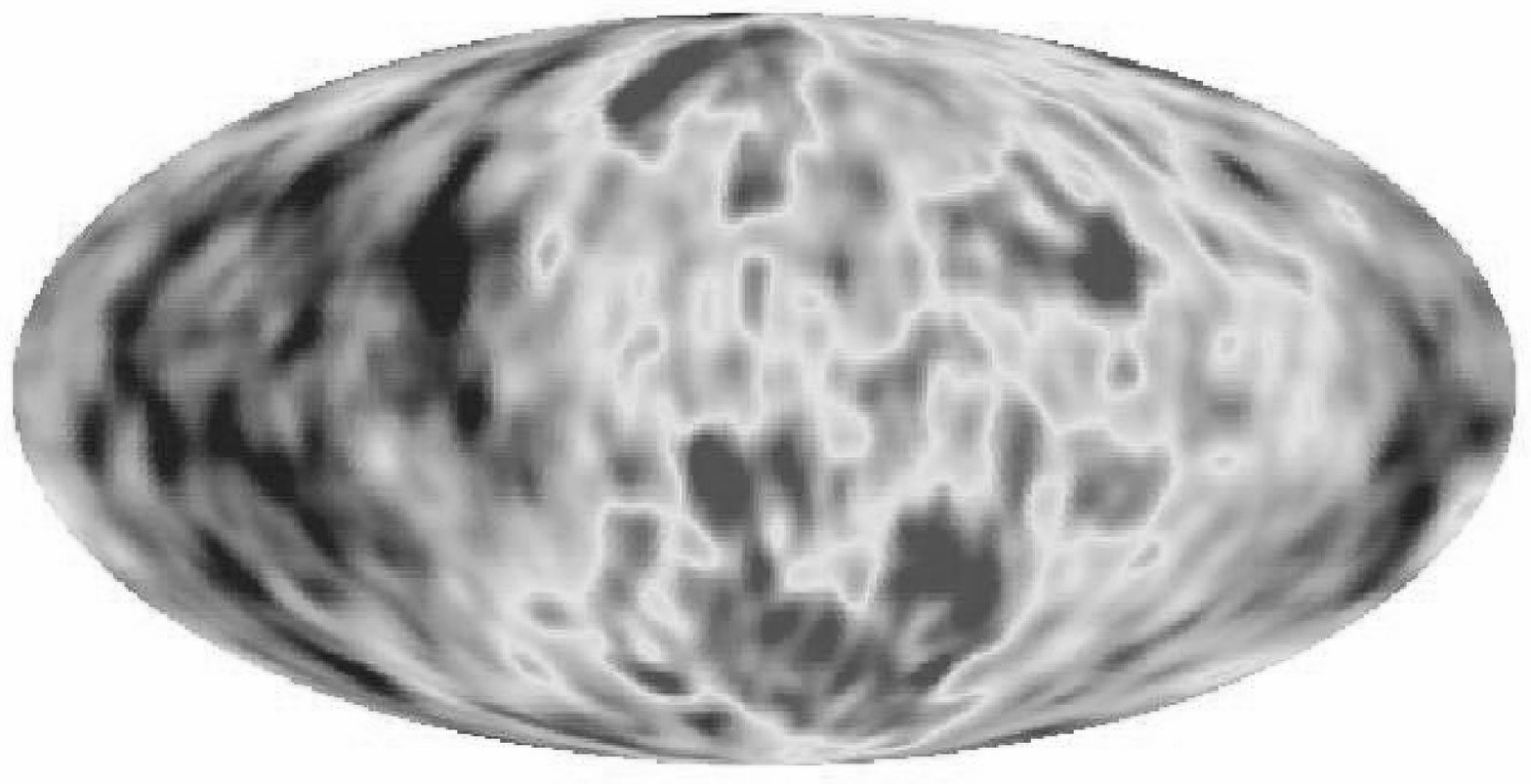}}
\resizebox{77mm}{!}{\includegraphics{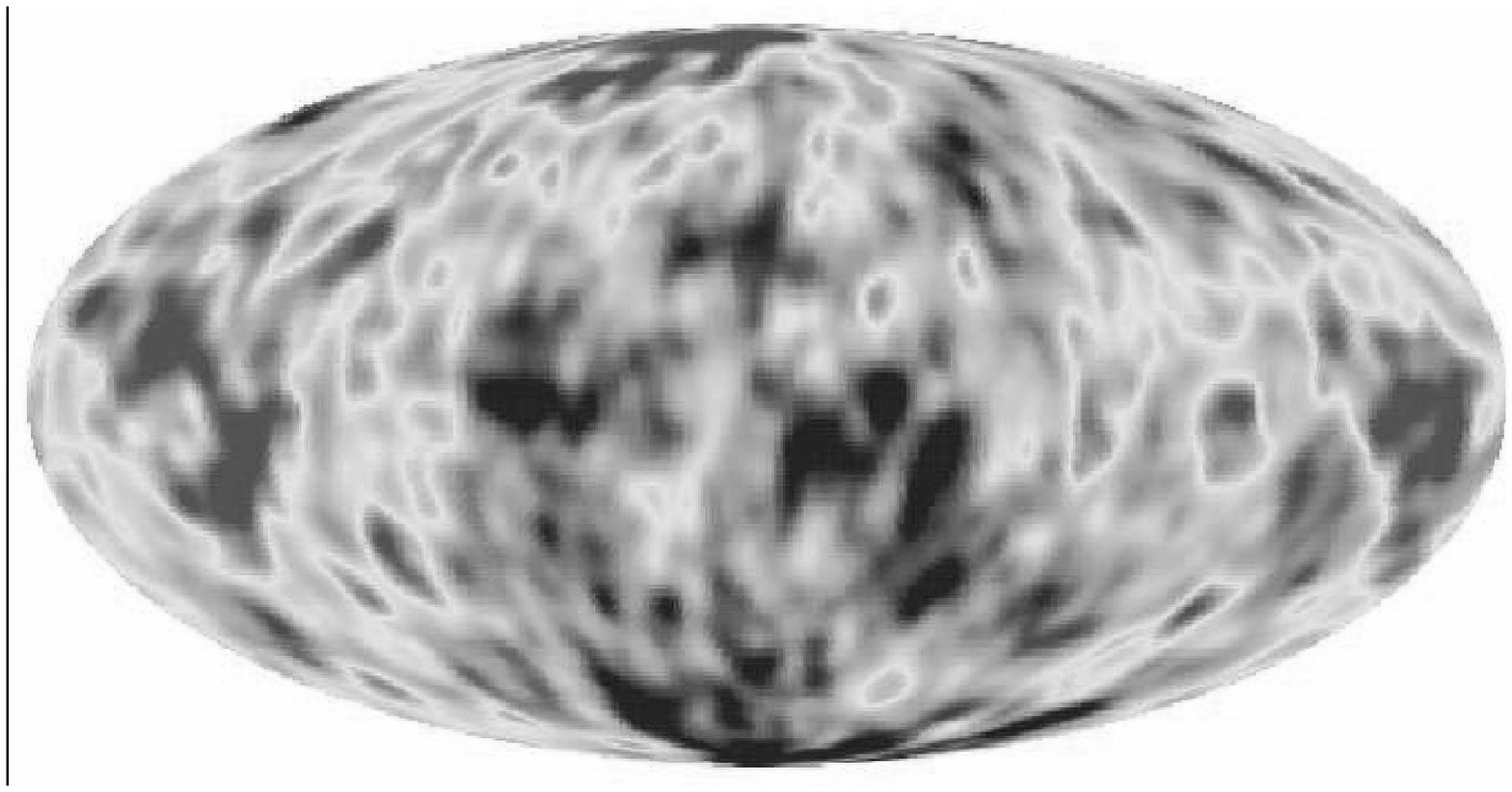}}
\caption{CMB maps for $\Omega_{\Lambda}=0$
(left) and,$\Omega_{\Lambda}=0.7$ (right).  The other parameters are as
given in the text.}\label{fig:lamaps}
\end{figure}

Our first results~\cite{lacmb} were all-sky CMB maps from which we
inferred the varying of the reduced linear energy density of the
strings $G\mu/c^2$ as a function of the cosmological constant by
normalizing the low-$\ell$ plateau of COBE.  Two such maps are shown
in Fig.~\ref{fig:lamaps}

Our result, obtained for
a flat FRW Universe with $h = 0.72$ and $\Omega_bh^2 = 0.002$ is:
\begin{equation}
\frac{G\mu}{c^2} = \left(0.695 + \frac{0.012}{1-\Omega_{\Lambda}}
\right)\times 10^{-6} \, .
\end{equation}
The cosmic string simulations all started at $\eta_0/4$ and ended at
$\eta_0$, the conformal time today.  This is a lower normalization
than obtained in Allen \textit{et al.}~\cite{allen96}, which is most
likely due to the more accurate incorporation of small-scale structure
in the string networks.

\subsection{Medium to Small Angle Maps}

\begin{figure}
\resizebox{52mm}{!}{\includegraphics{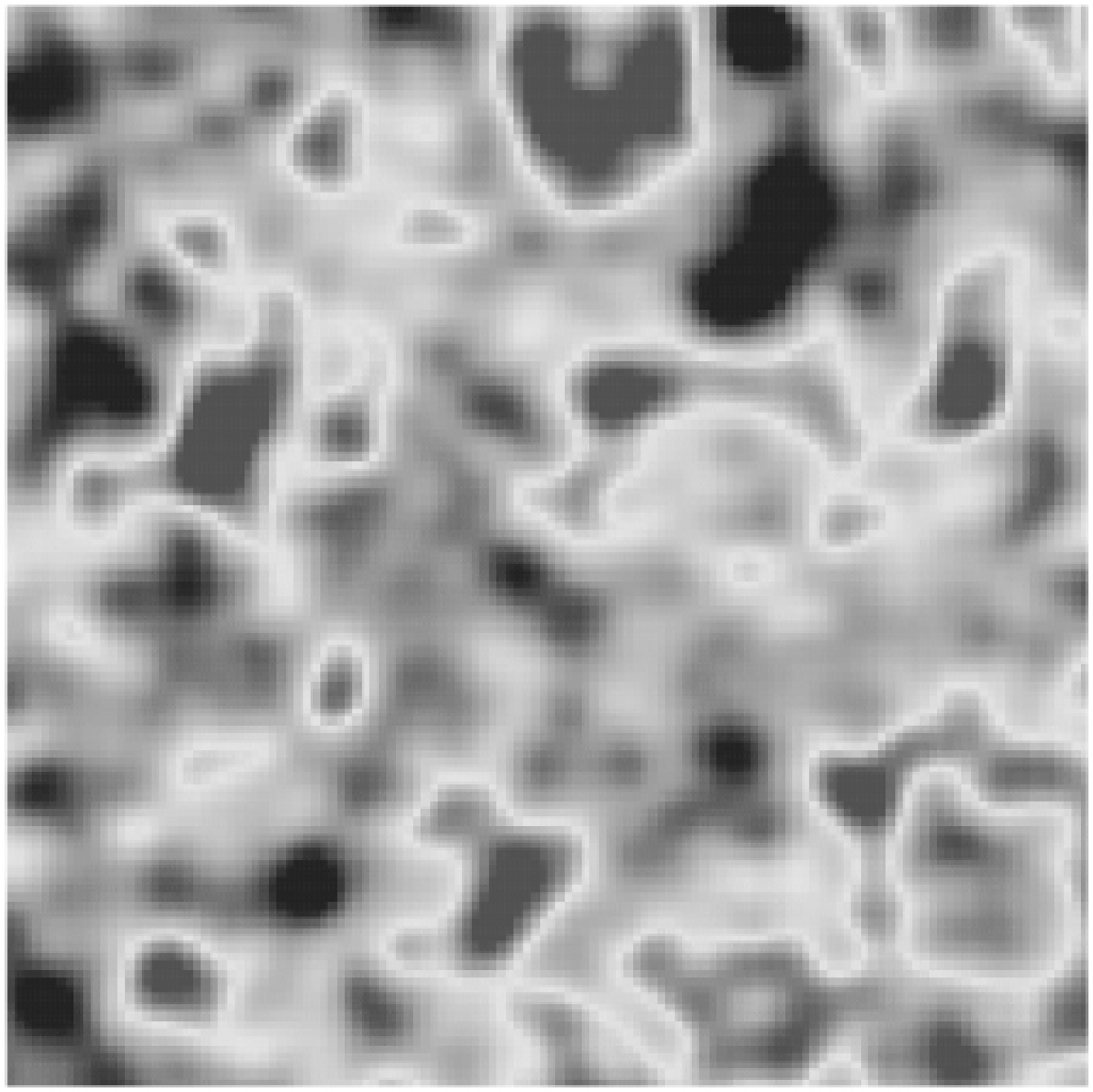}}
\resizebox{52mm}{!}{\includegraphics{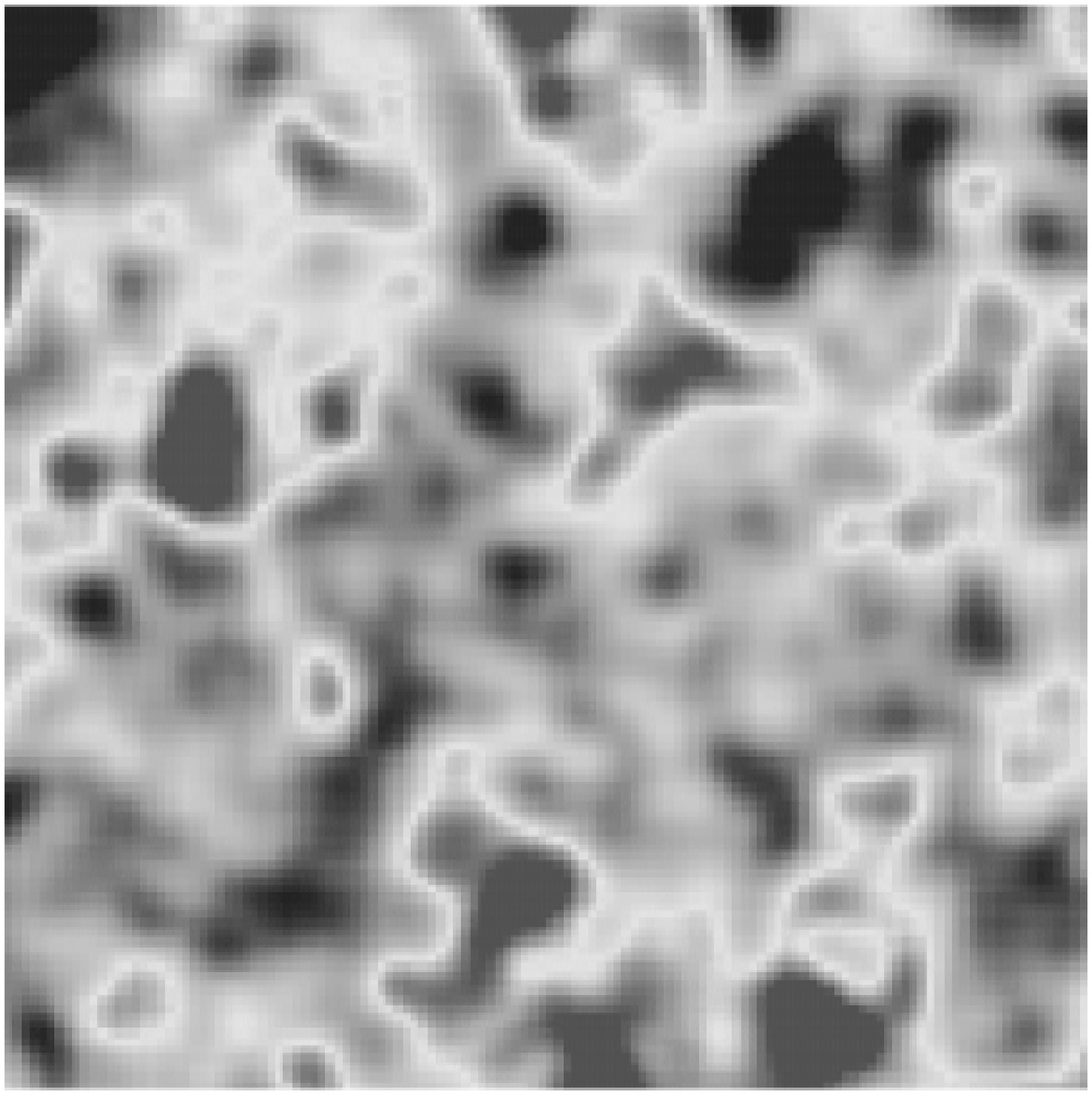}}
\resizebox{52mm}{!}{\includegraphics{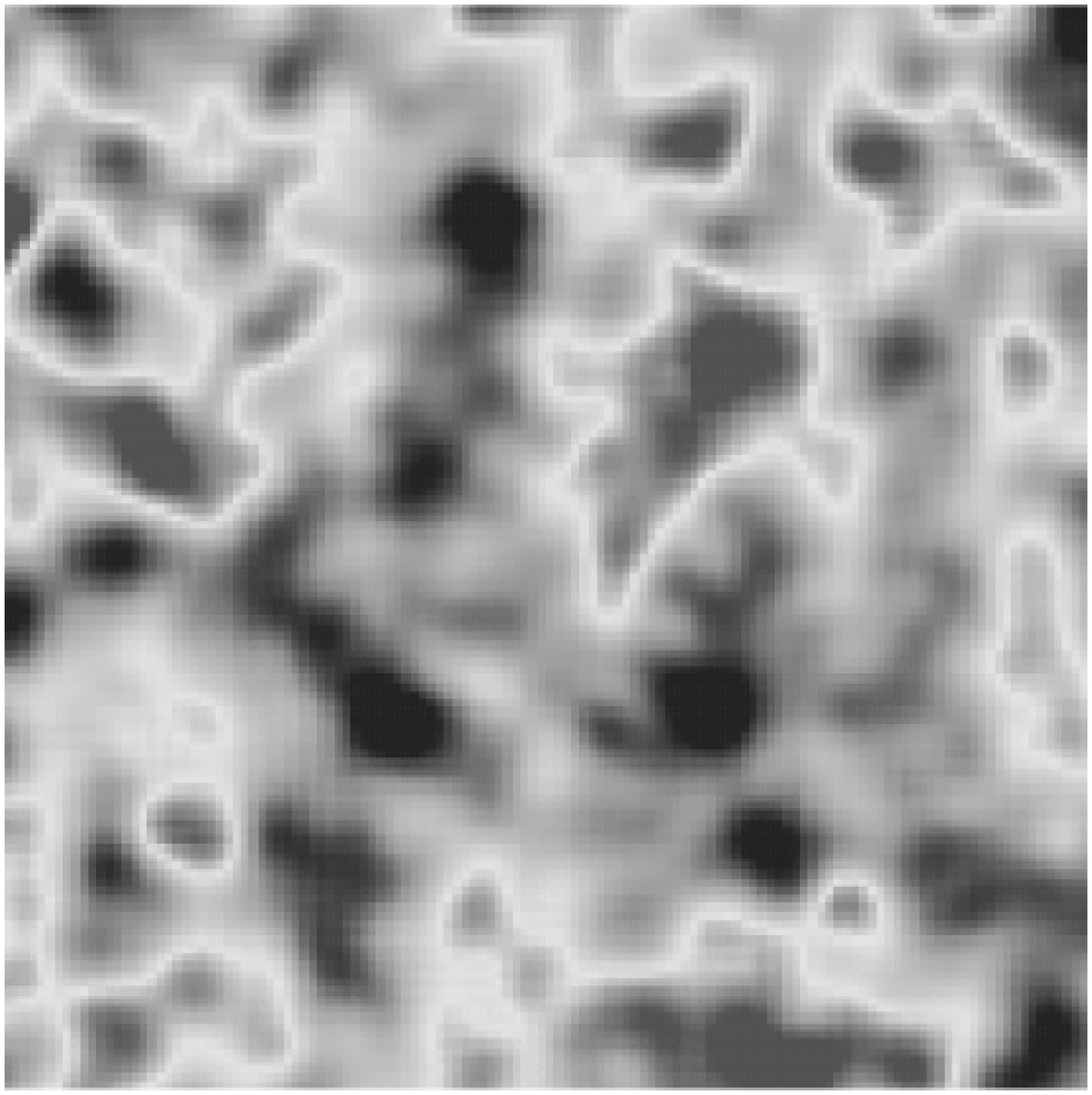}}
\caption{CMB maps of $6.4^{\circ}\times 6.4^{\circ}$,
$12.8^{\circ}\times 12.8^{\circ}$ and $25.6^{\circ}\times
25.6^{\circ}$.  The parameters are as given in the text.}\label{fig:maps}
\end{figure}

When studying smaller angular resolution, it is necessary to restrict
ourselves to patches of sky.  Fig.~\ref{fig:maps} shows three such
maps, each with 16384 square pixels computed for a flat FRW Universe
with the same parameters as for the all-sky maps, but with the
cosmological constant fixed at $\Omega_{\Lambda}=0.7$.  The
simulations span different epochs: $\eta_{rec}$ to $4\eta_{rec}$,
$2\eta_{rec}$ to $8\eta_{rec}$ and $4\eta_{rec}$ to $16\eta_{rec}$ for
the smallest to the largest maps.

\begin{figure}
\resizebox{78mm}{!}{\includegraphics{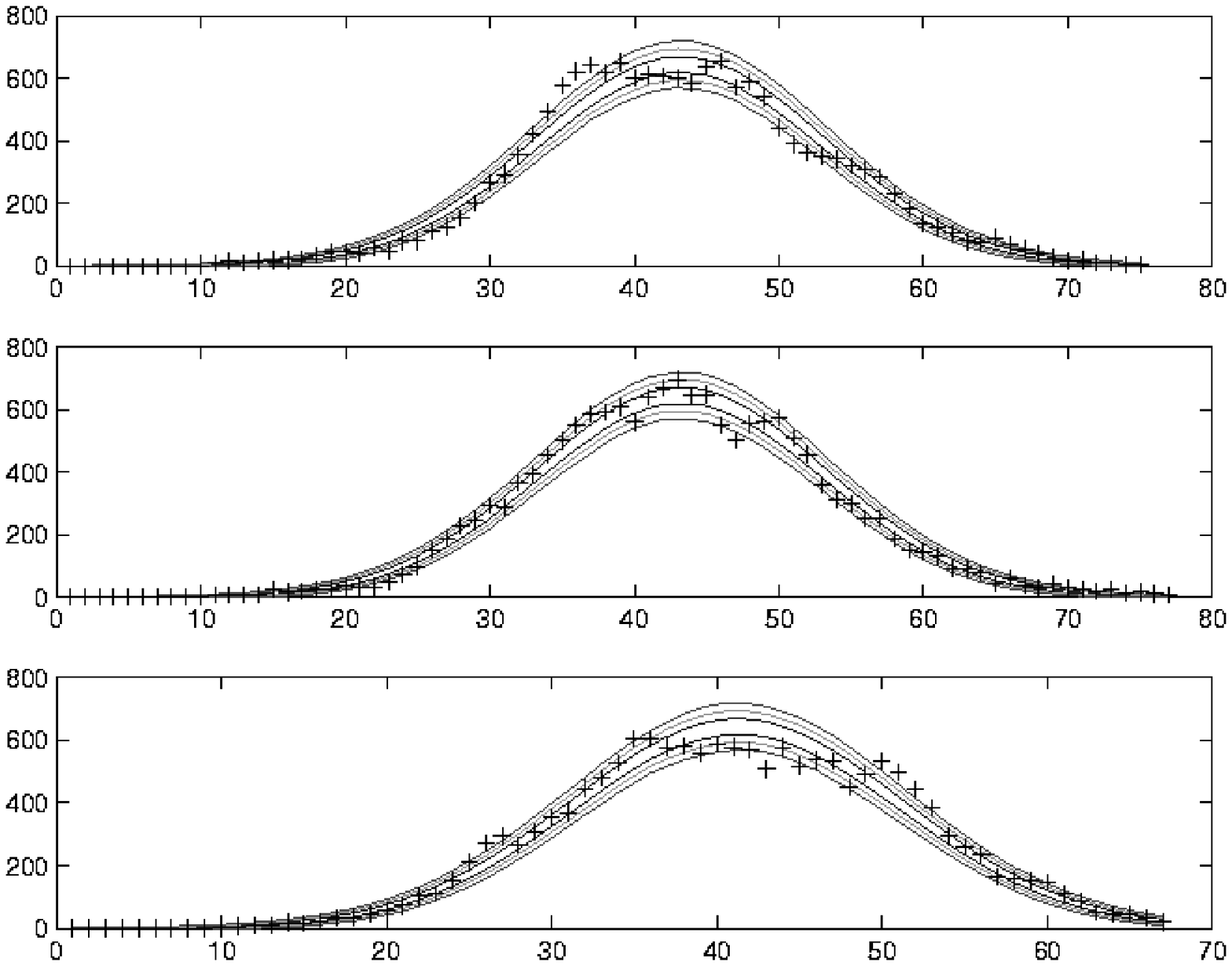}}
\resizebox{78mm}{!}{\includegraphics{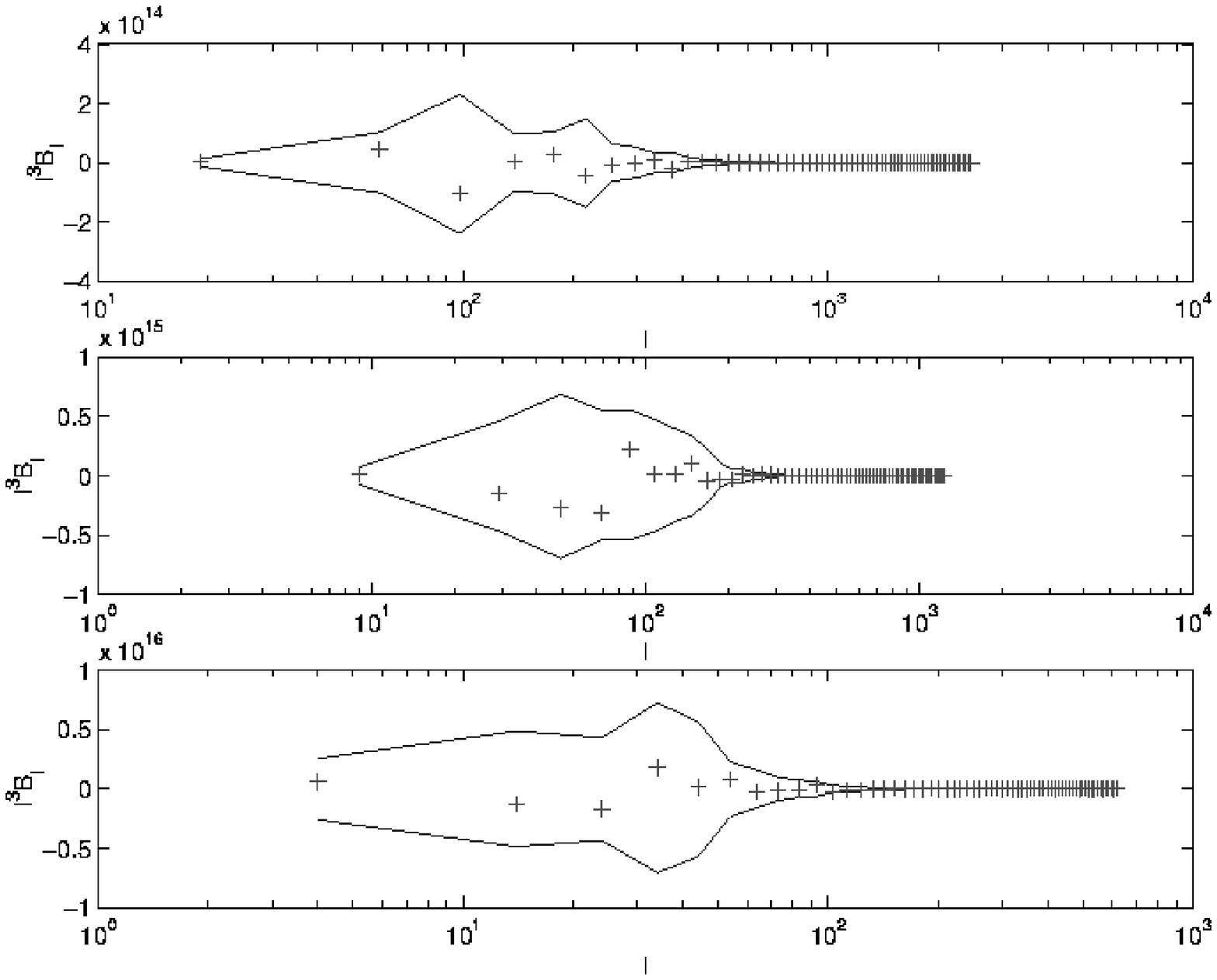}}
\caption{Pixel Temperature distribution for the maps shown in
Fig.~\ref{fig:maps} with the 1-$\sigma$, 2-$\sigma$ and 3-$\sigma$
contours of an ensemble of 1000 Gaussian maps (left) and the diagonal
bispectrum for the same maps with the 1-$\sigma$ contours the Gaussian
ensemble.}\label{fig:ng}
\end{figure}

We have performed a very basic search for non-Gaussianities and found
that the pixel temperature distribution is indeed non-Gaussian, but
that the completely diagonal bispectrum is not a good tool to extract
this signal.  This is most easily seen from Fig.~\ref{fig:ng}.  Note
that thesevery preliminary results come from simulations with a
limited dynamic range, a fact which tends to enhance the signatures of
individual strings.  Temperature contributions from the higher density
of strings at earlier times will tend to push the results towards
Gaussianity.  Further results will be published
elsewhere~\cite{hrcmb}.

\section{Non-Gaussian Inflation}

Departures from Gaussianity in inflationnary models can arise in a
number of circumstances, for example when there are many fields (e.g
Bartolo \textit{et al.}~\cite{bartolo2002}, Bernardeau \&
Uzan~\cite{bernardeau2002}) or when higher order perturbations are
taken into account (e.g. Acquaviva \textit{et
al.}~\cite{acquaviva2003}, Maldacena~\cite{maldacena2003}).  Other
authors have proposed methods to compute non-Gaussian inflationary
maps (e.g. Ligouri \textit{et al.}~\cite{liguori2003}).  Here, we
simply adapt the method outlined for cosmic strings (or other causal
sources) to inflation.  In this case, $q = 0$ ; hence, the integral in
Eq.~\ref{eq:soln} is zero as well, which reduces the computation by
more than half (there is further reduction in computing time from the
fact that the inverts of the fundamental matrices need not be
computed, but this is negligeable in comparison).  Hence, only the
initial conditions need to specified on a grid.  These can be obtained
from a field simulation of the inflaton~\cite{nginf}.

\section{Conclusion and Future Prospects}

In this talk, we have reviewed a method to compute maps of the CMB in
intrinsically non-Gaussian models.  We also presented preliminary
results on intermediary angle size CMB maps in the presence of cosmic
strings.  These maps exhibit some non-Gaussianities, but further
analysis may reveal they are not generic.  We are currently working on
smaller angle high resolution maps from cosmic strings~\cite{hrcmb}
and non-Gaussian inflation~\cite{nginf}.  Hopefully, these two very
different scenarios could be easily differentiated by their
non-Gaussian signature alone.

\section*{Acknoledgements}

All the simulations were performed on COSMOS IV, the Origin 3800
supercomputer, funded by SGI, HEFCE and PPARC.

\section*{References}
\bibliography{refs,myrefs,inf,wmap}

\end{document}